\newcommand{\beq}{\begin{equation}}
\newcommand{\eeq}{\end{equation}}

\documentclass[twocolumn, superscriptaddress, amsmath, amssymb, aps, prl, floatfix]{revtex4-2}

\usepackage{graphicx}
\usepackage{dcolumn}
\usepackage{bm}
\usepackage{amsmath}
\usepackage[colorlinks =true,
            linkcolor = blue,
            urlcolor  = blue,
            citecolor = blue,
            anchorcolor = blue]{hyperref}
\usepackage{amssymb}
    \usepackage{braket}
\usepackage{caption}
\usepackage{subcaption}
\usepackage{multirow}
\usepackage{color}

\definecolor{capri}{rgb}{0.0, 0.45, 0.73}

\usepackage{xcolor}

\begin{document}

\preprint{APS/123-QED}

\author{Zhifan Zhou}
\email{zhifanzhou12@gmail.com}
\affiliation{Joint Quantum Institute, University of Maryland and National Institute of Standards and Technology, College Park, Maryland 20742, USA}

\author{Yaxin Li}
\email{yli02@umd.edu}
\affiliation{Joint Quantum Institute, University of Maryland and National Institute of Standards and Technology, College Park, Maryland 20742, USA}


\title{Noncyclic geometric phase in three-level Ramsey interferometry for enhanced metrology}

\begin{abstract}

In a standard two-level Ramsey interferometer, the accumulated signal phase is linearly mapped to the readout phase. Here, we introduce three-level Ramsey interferometry, in which projected interference between internal paths reshapes this mapping through a noncyclic geometric phase response. Near a geodesic-closure transition, a small accumulated signal phase produces a sharply amplified readout-phase shift. We quantify the accompanying gain--visibility tradeoff and identify a finite operating window in which the amplified response can yield a net signal-to-noise-ratio gain in the presence of additional technical phase noise. A controllable Ramsey phase offset further positions this high-slope response at a chosen operating point, making the local enhancement accessible without requiring a large signal-induced phase. More broadly, these results establish a multilevel Ramsey route to enhanced phase sensitivity in quantum platforms, in which two signal-collecting internal paths interfere to produce a controllable noncyclic geometric response.

\end{abstract}
\maketitle

\begin{figure*}[t]
\begin{center}
\includegraphics[width=0.95\linewidth]{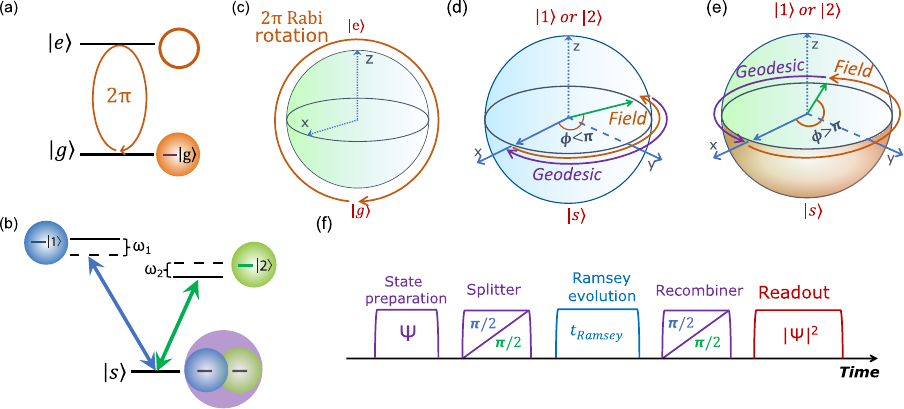}
\end{center}
\caption{\textbf{Geometric phase in a noncyclic three-level Ramsey interferometer.}
\textbf{(a,c)} A full \(2\pi\) Rabi cycle in a two-level system gives a global phase of \(\pi\), equal to half the enclosed solid angle on the Bloch sphere.
\textbf{(b)} V-type example: the shared state \(|s\rangle\) couples to two signal-collecting states \(|1\rangle\) and \(|2\rangle\) with transition frequencies \(\omega_1\) and \(\omega_2\), forming a projected internal-path interference signal.
\textbf{(d,e)} As the relative phase crosses \(\phi=\pi\), the geodesic closure switches branches, producing a sharp geometric-phase change in the projected readout phase.
\textbf{(f)} Ramsey sequence: preparation, splitting, evolution, recombination, and population readout.}
\label{fig-Scheme}
\end{figure*}

\paragraph{Introduction---}\hspace{-12pt}
Ramsey interferometry is a central method in quantum sensing and metrology\,\cite{Ramsey1950,DegenRMP2017}, with implementations across platforms including neutral atoms and trapped ions\,\cite{RamseyIon2021}, molecules\,\cite{Molecule2002}, and solid-state nitrogen-vacancy\,(NV) centers\,\cite{RMP_NV}. Standard two-level Ramsey interferometers, in which a signal phase accumulates linearly during the interrogation time, underlie applications ranging from precision magnetic-field sensing\,\cite{RMP_NV} to tests of gravitational redshift\,\cite{RMP15,Katori20} and searches for dark-matter signals through variations of fundamental constants\,\cite{Safronova18,Peik2023}. 
Ramsey sensitivity\,\cite{RMP15,RMP_NV} is typically improved by increasing the accumulated signal phase through longer coherence and interrogation times, reducing technical noise, controlling systematic shifts, and using quantum resources such as squeezing or entanglement\,\cite{RMPQuantum,pedrozo2020entanglement,Robinson2024SpinSqueezedClock}.
Complementary to these approaches, recent work has also improved metrological performance by optimizing Ramsey protocols, measurement strategies, and operating points\,\cite{kaubruegger2021quantum,Marciniak2022OptimalMetrology,overton2026adaptive}.

Here, we introduce a three-level Ramsey interferometer that reshapes the
signal-to-readout phase mapping through a noncyclic geometric phase response.
In this multilevel geometry, two signal-collecting internal paths are
coherently mapped onto a shared readout state, so that their amplitudes
interfere to define an effective readout phase. Near a geodesic-closure
transition\,\cite{samuel-bhandari,wagh1995measuring}, this readout phase can
acquire a sharply enhanced response to the accumulated signal phase. Within a
suitable operating window, after including the visibility-induced
projection-noise penalty, the enhanced slope can yield a net signal-to-noise
ratio\,(SNR) gain by reducing the contribution of additional technical noise
to the inferred signal-phase uncertainty. Furthermore, phase-offset control
positions this high-slope response at a desired operating point, allowing
small signal-induced phases to access the enhanced response. Because many
quantum sensors already provide more than two accessible internal
states\,\cite{DegenRMP2017}, this projected-path geometry offers a direct route
to geometric phase-response engineering in multilevel Ramsey platforms.

\begin{figure*}[t]
    \centering
    \includegraphics[width=0.79\textwidth]{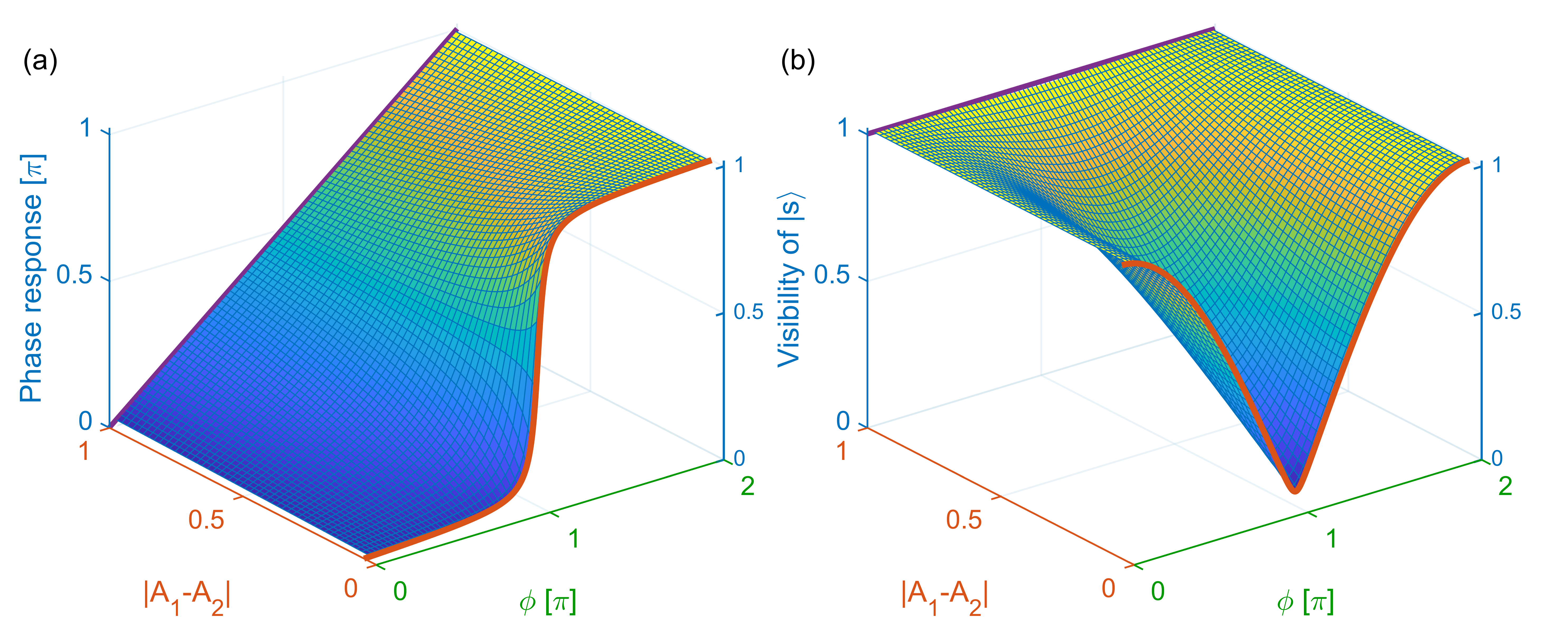}
\caption{\textbf{Geometric phase amplification and visibility tradeoff.}
\textbf{(a)} Phase-response landscape as a function of the accumulated differential phase \(\phi\) and the pathway imbalance \(|A_1-A_2|\).
The purple cut denotes the one-path limit \(|A_1-A_2|=1\), while the orange cut denotes a representative near-balanced geometric-response case with \(A_1=0.525\) and \(A_2=0.475\), corresponding to \(|A_1-A_2|=0.05\).
In the one-path limit, the projected internal two-path amplitude
reduces to $e^{i\phi/2}$, providing the reference slope $1/2$ used
to normalize the gain.
For nearly balanced pathways, the response becomes strongly nonlinear near the geodesic-closure transition at \(\phi=\pi\), producing an amplified local phase response.
\textbf{(b)} Corresponding projected visibility landscape, showing the visibility cost associated with the amplified response.
For the orange cut, the visibility minimum remains finite, \(V_{\min}=|A_1-A_2|=0.05\), while the maximum normalized phase-response gain is \(G_N^{\max}=1/|A_1-A_2|=20\).}
\label{fig-GPA}
\end{figure*}

\paragraph{Three-level noncyclic geometric phase---}\hspace{-12pt}
To illustrate how the internal interference acquires a geometric character, we first recall a familiar two-level example: a full $2\pi$ Rabi cycle returns the state to itself with a global $\pi$ phase, equal to half the solid angle enclosed on the Bloch sphere\,\cite{spinHalf,rauch1975verification,FastGates}\,(Fig.\,\ref{fig-Scheme}(a,\,c)).
In the three-level Ramsey interferometer\,(Fig.\,\ref{fig-Scheme}(b,\,d,\,e)), the geometric phase arises from a noncyclic trajectory associated with projected internal-path interference.
On the Bloch-sphere representation, the two signal-collecting components define an open trajectory whose geodesic closure depends on the relative rotation angle\,\cite{berry1984quantal,pancharatnam1956generalized,
samuel-bhandari,bhandari19912,GeoPhaseJump,zhou2025geometric}.
As the relative rotation crosses $\phi=\pi$, the geodesic closure switches branches, producing a noncyclic geometric phase jump.
An initial Ramsey phase offset shifts the position of this branch-switching region, creating a controllable operating point where a small phase shift is mapped into a much larger readout-phase change.

Multilevel Ramsey sensing has also been explored\,\cite{PRB2016,NPJqi2018,
Madasu2024GeometricRamsey,TPS_PRL}, including three-level NV-center schemes
for zero- or near-zero-field sensing\,\cite{DU2018,ThreeLevelControlPRL2021,
Genko2022,DuPRApplied2024}. These works illustrate the utility of multilevel Ramsey structure across a range of control and readout geometries. Here, we focus on a different geometry: two signal-collecting components are combined through a shared-state projected readout, producing a noncyclic geometric phase response in the signal-to-readout phase mapping.
The scheme can be realized in three-level systems where two
coherently driven transitions share a common state, including neutral atoms and
solid-state spin systems\,\cite{Safronova18,RMP_NV,DU2018,
ThreeLevelControlPRL2021,Genko2022,DuPRApplied2024}. The V-type schematic in
Fig.\,\ref{fig-Scheme} is one representative level ordering; the projected-interference description can apply to ladder and \(\Lambda\)-type configurations when the pulse sequence creates two phase-coherent signal-accumulating components projected onto a common readout state.


The Ramsey sequence uses a splitter, signal evolution, and a recombiner
formed by nominal \(\pi/2\) rotations on the two transitions connecting
the shared state to the signal-collecting states
\,(Fig.\,\ref{fig-Scheme}(b,\,f)). After the final readout pulses, the shared-state signal is determined by the interference of two Ramsey components with relative population weights \(A_1\) and
\(A_2\). The corresponding projected complex amplitude can be written as
\begin{equation}
A_1 e^{i\phi/2}+A_2 e^{-i\phi/2}
=
\cos\left(\frac{\phi}{2}\right)
+i\epsilon\sin\left(\frac{\phi}{2}\right),
\label{eq:complex_amplitude}
\end{equation}
where \(A_1+A_2=1\), \(\epsilon=A_1-A_2\), and \(\phi\) is the accumulated relative signal phase between the two Ramsey components. The visibility and projected readout phase are then given by the modulus and argument of the complex amplitude in Eq.\,\eqref{eq:complex_amplitude},
 \begin{align}
V_s(\phi)
&=
\sqrt{
\cos^2\!\left(\frac{\phi}{2}\right)
+
\epsilon^2
\sin^2\!\left(\frac{\phi}{2}\right)
},
\label{eq:envelop}\\
\Phi(\phi)
&=
\arctan
\left[
\frac{
\epsilon \sin\!\left(\phi/2\right)
}{
\cos\!\left(\phi/2\right)
}
\right].
\label{eq:phase}
\end{align}
Here $\Phi(\phi)$ is taken continuously across $\phi=\pi$. For nearly balanced pathways, Eqs.\,(2) and (3) show a critical region near
$\phi=\pi$, where the visibility minimum and sharp projected-phase variation
reflect the noncyclic geometric phase jump at geodesic
closure\,\cite{wagh1995measuring,NeutronNoncyclic,Sjoqvist2001Noncyclic,Xiao2010RMP,vanDijk2010OE,Rantaeskola2025Optica,Gil2026OL}.

\begin{figure*}[t]
\centering
\includegraphics[width=0.96\linewidth]{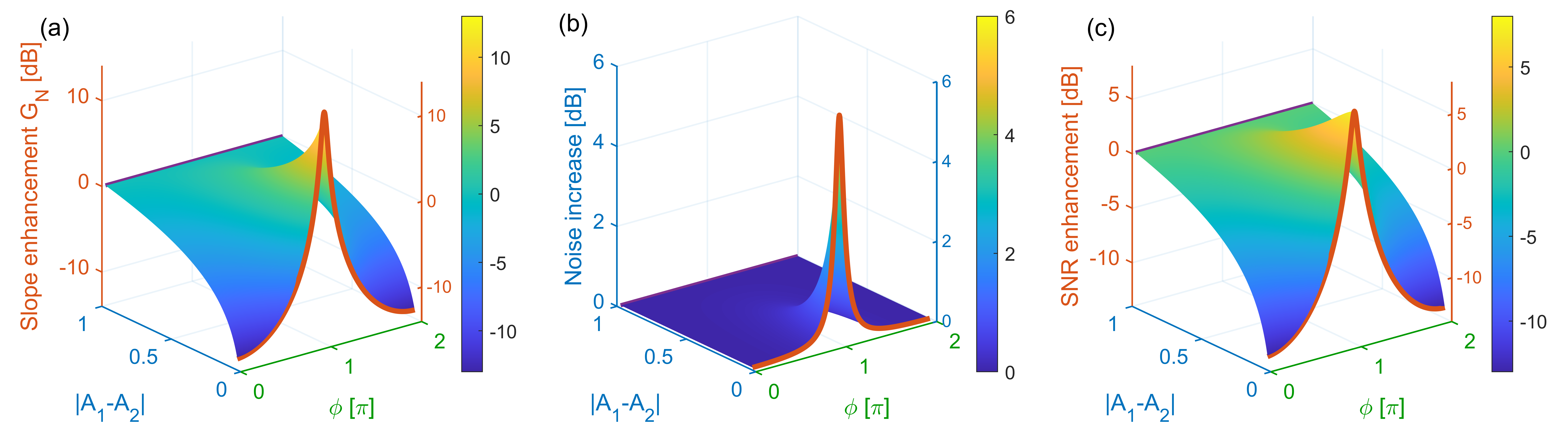}
\caption{
\textbf{Sensitivity landscape of the noncyclic geometric phase response.}
\textbf{(a)} Normalized phase-response gain \(G_N\), \textbf{(b)} visibility-induced phase-noise increase, and \textbf{(c)} resulting net SNR enhancement as
functions of the accumulated phase \(\phi\) and pathway imbalance
\(|A_1-A_2|\). The purple and orange cuts denote the one-path and representative near-balanced cases, respectively. For the orange cut, the maximum normalized
phase-response gain is \(G_N^{\max}=1/|A_1-A_2|=20\), while the minimum
visibility remains finite, \(V_{\min}=|A_1-A_2|=0.05\). The positive SNR-enhancement region in \textbf{(c)}, calculated for \(N=3\times10^3\)
and \(\xi_{\rm tech}=0.1\,{\rm rad}\), identifies finite operating windows
where, after including the visibility-induced projection-noise penalty, the
amplified phase response reduces the input-referred contribution of additional
technical phase noise.
}
\label{fig-Sensitivity}
\end{figure*}

\paragraph*{Geometric phase amplification and visibility tradeoff---}\hspace{-12pt}
Figure\,\ref{fig-GPA}  visualizes the resulting gain--visibility tradeoff as a
function of $|\epsilon|$. The horizontal axis is the accumulated
differential phase $\phi$, while the plotted phase response is the
readout phase $\Phi(\phi)$. Throughout, we characterize the phase of the projected two-path
envelope defined by Eq.\,(\ref{eq:complex_amplitude}), with the common phase factor removed so
that the two components carry phases $+\phi/2$ and $-\phi/2$. In the one-path limit, the projected internal two-path amplitude reduces to
$e^{i\phi/2}$, providing the reference slope $1/2$ used to normalize the
gain below. The plotted phase response spans $0$ to $\pi$ as $\phi$ changes from
$0$ to $2\pi$. The orange cut marks the representative near-symmetric amplified case
$|\epsilon|=0.05$, while the purple cut marks the one-path limit $|\epsilon|=1$.

Near the geodesic-closure transition, the local phase-response slope
$|\partial\Phi/\partial\phi|$ reaches $1/(2|\epsilon|)$, while the
minimum projected visibility is $V_{\min}=|\epsilon|$. Since the
reference Ramsey response has slope $1/2$ in this phase convention,
we define the normalized local phase-response gain $G_N(\phi)$
relative to this benchmark. For the present projected response,
$G_N(\phi)=|\epsilon|/V_s^2(\phi)$ (see Supplemental
Material\,\cite{SM}), which reaches $G_N^{\max}=1/|\epsilon|$.

\begin{figure*}[t]
  \centering
  \includegraphics[width=0.98\textwidth]{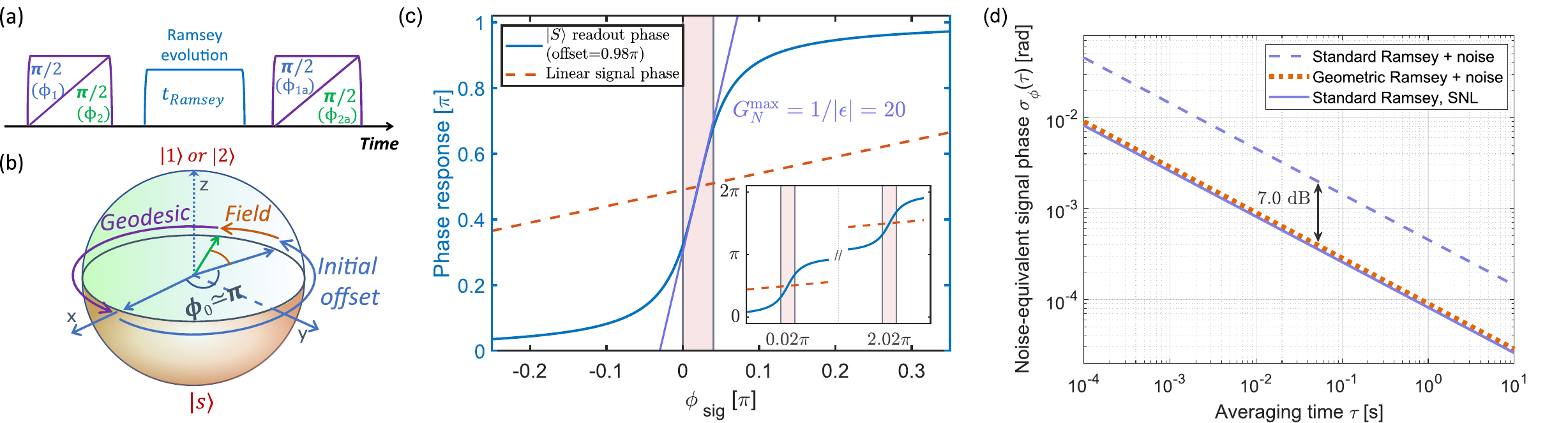}
\caption{\textbf{Offset-controlled geometric response and projected phase sensitivity.}
\textbf{(a)} Offset-controlled Ramsey protocol. The first Ramsey pulse sets
the initial relative offset phase \(\phi_{\rm off}=\phi_1-\phi_2\), and the
final Ramsey pulse reads out the projected phase through the scanned phases
\(\phi_{1a}\) and \(\phi_{2a}\).
\textbf{(b)} Bloch-sphere illustration of the offset-controlled trajectory:
the initial offset places the open trajectory near the geodesic-closure branch
region before signal-induced evolution.
\textbf{(c)} Offset-displaced phase response for
\(|A_1-A_2|=0.05\) and \(\phi_{\rm off}=0.98\pi\).
The shaded region marks the finite high-slope window,
\(\phi_{\rm sig}/\pi\in[0,0.04]\), in which the signal phase is mapped
near the phase-response jump. The maximum normalized gain relative to
the one-path Ramsey reference slope is
\(G_N^{\max}=1/|\epsilon|=20\).
\textbf{(d)} Projected noise-equivalent signal phase under repeated
finite-window measurements. The curves use \(N=3\times10^3\), a common cycle
time \(T_c=20\,\mu{\rm s}\), and additional technical phase noise
\(\xi_{\rm tech}=0.1\,{\rm rad}\) for the noisy cases. The geometric curve uses the finite-window effective enhancement over the shaded region in \textbf{(c)}, including the enhanced local response and visibility penalty. All curves follow the expected \(\tau^{-1/2}\) averaging; under the same
sampling cycle and technical-noise level, the offset-tuned geometric response
improves the plotted noise-equivalent phase sensitivity by about \(7\,{\rm dB}\)
relative to a noisy standard Ramsey, using
\(10\log_{10}(\sigma_{\rm std}/\sigma_{\rm geo})\).}
\label{fig-Shortcut}
\end{figure*}

\paragraph*{Sensitivity window---}\hspace{-12pt}
The amplified noncyclic slope improves metrology only within an operating
window where the enhanced phase response remains useful after the
visibility-induced projection-noise penalty is included. The projection-noise
contribution is $\Delta\Phi_{\rm PN}=1/[V_s(\phi)\sqrt{N}]$, while
$\xi_{\rm tech}$ denotes additional technical phase noise. We model the total
phase uncertainty as
$\Delta\Phi^2=(\Delta\Phi_{\rm PN})^2+\xi_{\rm tech}^2$.
For the sensitivity comparison, we express input-referred
uncertainties in the same one-path Ramsey phase convention
and denote the resulting Ramsey-equivalent phase uncertainty
by $\Delta\phi$.
Local error propagation then gives
\begin{equation}
\Delta \phi
=
\frac{\Delta\Phi}{G_N(\phi)}
=
\frac{1}{G_N(\phi)}
\sqrt{
\frac{1}{V_s^2(\phi)N}
+
\xi_{\rm tech}^2
},
\label{eq:inferred_phase_uncertainty}
\end{equation}
where $N$ is the atom number. 

Equation\,\eqref{eq:inferred_phase_uncertainty} captures the local tradeoff at fixed internal-path
imbalance: the enhanced gain $G_N(\phi)$ reduces the inferred
signal-phase uncertainty, while reduced visibility increases the
projection-noise contribution through $1/V_s(\phi)$. At the
shot-noise limit\,(SNL), the gain--visibility balance leaves the
optimal single-shot phase sensitivity unchanged. With additional
technical phase noise, the gain can produce a finite operating window
with net SNR enhancement. For the numerical sensitivity landscape in Fig.\,\ref{fig-Sensitivity}, we use a representative atom
number $N=3\times10^3$ and $\xi_{\rm tech}=0.1\,\mathrm{rad}$, for which the
baseline SNL phase uncertainty $1/\sqrt{N}\simeq0.018~\mathrm{rad}$ is
smaller than $\xi_{\rm tech}$. Fig.\,\ref{fig-Sensitivity} maps this operating region through
the local gain in (a), the visibility penalty in (b), and the resulting
net SNR enhancement in (c). Along the representative near-balanced
cut \(|A_1-A_2|=0.05\), the local net SNR enhancement reaches about
\(7.3\,{\rm dB}\).

\paragraph*{Phase-offset control of the geometric response---}\hspace{-12pt}
The geometric enhancement is a local phase-response advantage within a finite operating window. To position this high-slope response at a desired operating point without requiring the signal itself to accumulate the corresponding bias phase, we introduce an initial relative Ramsey phase offset. This offset places the interferometer near the geodesic-closure branch region associated with the noncyclic geometric phase response\,\cite{samuel-bhandari,wagh1995measuring}. The offset-controlled protocol and its geometric origin are illustrated in Fig.\,\ref{fig-Shortcut}(a,b).

The preset offset, $\phi_{\rm off}=\phi_1-\phi_2$, biases the interferometer toward the branch region before the signal is applied. For a differential shift, the relevant input is the final accumulated signal
phase \(\phi_{\rm sig}\), which reduces to
\(\phi_{\rm sig}=\Delta\omega T_{\rm sig}\) for a constant shift over a signal
window \(T_{\rm sig}\). Thus, the operating window is defined in accumulated
phase rather than in measurement time; the corresponding time scale depends
on how the phase is generated. In the Bloch-sphere representation, the offset sets the starting point of the open trajectory relative to the branch region, so that small signal-induced phase changes are measured within the high-slope geometric-response window.


Figure\,\ref{fig-Shortcut}(c) shows the corresponding offset-displaced phase response. With $\phi_{\rm off}=0.98\pi$, the high-slope noncyclic response is reached by a small additional signal phase, such that $\phi_{\rm tot}=\phi_{\rm off}+\phi_{\rm sig}\simeq\pi$. Thus, the near-$\pi$ bias phase needed to reach the branch region is supplied by the Ramsey-pulse phases rather than by signal accumulation during free evolution, while the useful signal phase is measured locally within the high-slope window.

To connect this local phase response to the projected noise-equivalent signal phase in Fig.\,\ref{fig-Shortcut}(d), we use the single-shot input-referred signal-phase uncertainty $\Delta\phi$ defined by the local error-propagation model above. For independent samples with cycle time $T_c$, averaging over a total time $\tau$ gives
\begin{equation}
\sigma_\phi(\tau)=\Delta\phi\sqrt{\frac{T_c}{\tau}} .
\label{eq:phase_projection}
\end{equation}
This quantity represents the input-referred signal-phase noise after repeated finite-window measurements.

As a representative finite-window phase projection, we use the same
\(N=3\times10^3\) and \(\xi_{\rm tech}=0.1\,{\rm rad}\) as in Fig.\,\ref{fig-Sensitivity},
and introduce a cycle time \(T_c=20\,\mu{\rm s}\), motivated by
microsecond-to-tens-of-microseconds timescales in pulsed Rydberg
platforms\,\cite{Bohaichuk2022RydbergTransient,Kurdak2025RydbergDressing}.
Here, \(\xi_{\rm tech}=0.1\,{\rm rad}\) is comparable to the finite signal-phase
window in Fig.\,\ref{fig-Shortcut}(c), whose upper edge is
\(0.04\pi\simeq0.126\,{\rm rad}\), so the noisy standard-Ramsey single-shot
SNR remains of order unity. The geometric curve is evaluated over the offset-shifted operating window
in Fig.~4(c), with $\phi_{\rm off}=0.98\pi$ and $|\epsilon|=0.05$,
including both the enhanced local phase-response gain and the visibility
penalty, with the finite-window averaging procedure detailed in the
Supplemental Material\,\cite{SM}. This finite-window value should be distinguished from the peak local SNR
enhancement of about \(7.3\,{\rm dB}\) in Fig.\,\ref{fig-Sensitivity}(c):
here the plotted improvement uses the finite-window effective enhancement over the offset-shifted high-slope window. As shown in
Fig.\,\ref{fig-Shortcut}(d), all curves follow the expected
\(\tau^{-1/2}\) averaging behavior, while the offset-tuned geometric response
improves the plotted noise-equivalent signal-phase sensitivity by about
\(7\,{\rm dB}\), corresponding to about a factor-of-five reduction relative to
noisy standard Ramsey under the same sampling cycle and technical-noise level.

The same input-referred-noise reduction can also apply to slowly varying
technical phase noise that does not average down as white noise\,
\cite{DegenRMP2017,RMP15}.
If a residual technical phase-noise component $\xi_{\rm slow}$ produces a phase
floor in standard Ramsey, the geometric response reduces its input-referred
contribution to approximately $\xi_{\rm slow}/G_N$ within the same local
operating range. Thus, beyond the white-noise projection shown in
Fig.\,\ref{fig-Shortcut}(d), the enhanced local gain can also lower technical
phase floors associated with analyzer-phase drift, detection offsets, or other
slowly varying imperfections.

A related practical consideration is sensor number, or atom number in atomic
implementations. As Eq.\,\eqref{eq:inferred_phase_uncertainty} shows,
increasing $N$ reduces the projection-noise contribution and improves the
absolute sensitivity until the total uncertainty approaches the technical-noise
floor set by $\xi_{\rm tech}$. The geometric gain lowers the input-referred
contribution of this technical noise, thereby extending the
projection-noise-dominated $N^{-1/2}$ improvement to larger sensor numbers.
In real sensors, larger $N$ can also introduce or enhance technical-noise
contributions through density-dependent shifts, collisions, loading
fluctuations, and detection noise\,\cite{swallows2011suppression,szmuk2015stability}. For technical-noise
contributions that can be controlled or calibrated, this reduction can help
larger sensor numbers remain useful over a wider operating range.

\paragraph*{Discussion---}\hspace{-12pt}
The gain--visibility tradeoff also admits a useful analogy to dark-fringe
operation in interferometric metrology\,\cite{Bond2017Interferometer}.
Near a dark fringe, small phase perturbations are detected through departures
from destructive interference, at the cost of reduced detected signal and
increased projection or shot noise. In the present three-level Ramsey
geometry, an analogous tradeoff emerges near the geodesic-closure transition:
destructive interference between the two projected internal paths reduces the
visibility while the noncyclic geometric phase response produces a large local
readout-phase slope. The dark-fringe analogy therefore provides a simple
physical picture of the central gain--visibility tradeoff.

Because this cost arises from projection noise, the dark-fringe perspective
also suggests a natural connection to quantum-noise reduction.
Nonclassical resources such as spin squeezing or entanglement could further
reduce the projection-noise cost\,\cite{RMPQuantum,pedrozo2020entanglement,
Robinson2024SpinSqueezedClock}, analogous in spirit to squeezed-state
enhancement of interferometers\,\cite{Caves1981,Schnabel2017}.


\paragraph*{Conclusion---}\hspace{-12pt}
In summary, we have proposed a three-level Ramsey interferometer in which coherent interference between two signal-sensitive internal pathways gives rise to a noncyclic geometric phase response. This enhanced phase response can reduce the inferred contribution of additional technical noise and improve the useful SNR within a local operating window. Phase-offset control positions this high-slope noncyclic response at a chosen operating point, making the enhancement accessible to small signal-induced phase changes without requiring a large accumulated signal phase. Together, these elements establish a Ramsey-compatible multilevel sensing strategy in which internal-path interference converts the accumulated signal phase into a controllable geometric enhancement of the local phase response.


\begin{acknowledgments}
We thank Steven L. Rolston for helpful discussions, insightful comments on the manuscript, and support during the development of this work. We also thank J. V. Porto, Paul D. Lett, William D. Phillips, and Sebastian Osorio Perez for valuable discussions and technical assistance. Z.\,Z. was supported by the Air Force Office of Scientific Research under Grant No. FA9550-23-1-0039. Y.\,L. was supported by MAQP Grant No. W911NF-24-2-0107.
\end{acknowledgments}

\bibliography{sample}

\newpage
\clearpage
\onecolumngrid
\centerline{Supplementary material for}
\bigskip
\bigskip

\centerline{\bf Noncyclic geometric phase in three-level Ramsey interferometry for enhanced metrology}

\newcommand{\Sref}[2]{\hyperlink{Sref:#1}{[#2]}}

\renewcommand{\theequation}{S\arabic{equation}}
\renewcommand{\thefigure}{S\arabic{figure}}
\setcounter{figure}{0}
\setcounter{equation}{0}
\setcounter{page}{1}

\bigskip

\centerline{Zhifan Zhou and Yaxin Li}

\centerline{Joint Quantum Institute, University of Maryland and National Institute}
\centerline{of Standards and Technology, College Park,
Maryland 20742, USA}

\section{Branch-resolved population view of the three-level Ramsey response}

The main text formulates the three-level Ramsey response in terms of an effective visibility and a projected readout phase. Here we give a branch-resolved population-level representation that illustrates how two single-frequency Ramsey components combine into the projected shared-state fringe. This construction is intended as a transparent population-level decomposition of the projected signal; the effective complex-amplitude formulation in the next section gives the corresponding visibility and phase response used in the main text.

\begin{figure}[h]
    \centering
    \includegraphics[width=0.98\textwidth]{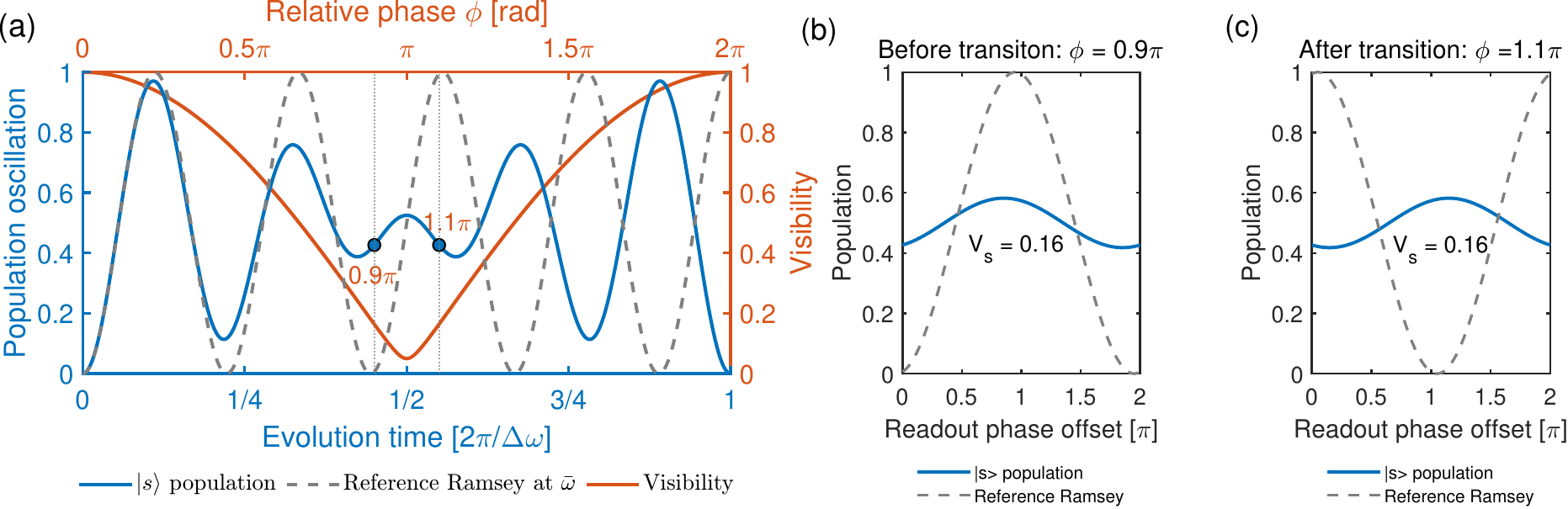}
 \caption{\textbf{Population oscillation and visibility near the noncyclic
transition.} (a) Projected shared-state population $P_s(t)$ as the relative
phase $\phi=(\omega_1-\omega_2)t$ evolves. The orange curve shows the visibility envelope $V_s(\phi)$ for
$|\epsilon|=|A_1-A_2|=0.05$, as defined in the main text, and the
dashed gray curve shows a reference Ramsey curve at the average
frequency $\bar{\omega}$, used only as a common phase guide. The marked points indicate
$\phi=0.9\pi$ and $\phi=1.1\pi$, on opposite sides of the visibility minimum
near $\phi=\pi$. (b,c) Local Ramsey fringes obtained by scanning the readout
phase offset at the two marked operating points. The projected $|s\rangle$
population fringe has finite visibility away from $\phi=\pi$ and is shifted
relative to this reference phase guide.}
    \label{fig:S1_population_visibility}
\end{figure}

We model the readout as a projected two-channel Ramsey signal, in which
the two signal-state components are resolved as distinguishable population
channels. This can be realized, for example, by state-selective addressing, or an equivalent reconstruction
of the projected signal. Denoting the two signal states by \(|1\rangle\)
and \(|2\rangle\), the corresponding single-channel Ramsey populations are
\begin{equation}
P_j(t)
=
\frac{A_j}{2}\left[1+\cos(\omega_j t)\right],
\qquad j=1,2,
\end{equation}
where $A_1+A_2=1$ specifies the relative pathway weights. Since the total population is conserved, the shared-state population is the complement of the total signal-state population,
\begin{equation}
P_s(t)
=
1-P_1(t)-P_2(t).
\end{equation}
Equivalently,
\begin{equation}
P_s(t)
=
\frac{1}{2}
\left[
A_1\left(1-\cos\omega_1 t\right)
+
A_2\left(1-\cos\omega_2 t\right)
\right].
\label{eq:SM_population_direct}
\end{equation}
This expression shows that the measured $|s\rangle$ population is obtained from the complement of the weighted sum of two single-frequency Ramsey components. As the relative phase $\phi=(\omega_1-\omega_2)t$ approaches $\pi$, the projected oscillating term in $P_s(t)$ is strongly suppressed, producing the visibility minimum shown in Fig.\,\ref{fig:S1_population_visibility}(a).

The marked points in Fig.\,\ref{fig:S1_population_visibility}(a) illustrate
the local Ramsey response on the two sides of this minimum. In
Figs.\,\ref{fig:S1_population_visibility}(b) and
\ref{fig:S1_population_visibility}(c), the relative phase is fixed at
\(\phi=0.9\pi\) or \(1.1\pi\), while an additional readout phase offset is
scanned. The projected \(|s\rangle\) population then appears as a Ramsey
fringe with reduced but finite contrast and with a shifted phase relative to
the reference phase guide at \(\bar{\omega}\). This shift is already visible
from the population response itself: as the system crosses the visibility
minimum, the fitted readout fringe changes branch. The signal phase \(\phi\) varies smoothly, while the branch change appears in the projected readout phase extracted from the population fringe.

The following section makes this projected-channel population response quantitative using the effective visibility and projected phase introduced in the main text.

\section{Effective visibility and projected phase}
\label{sec:SM_effective_fringe}


Starting from the population-level expression in Eq.\,\eqref{eq:SM_population_direct}, the oscillatory part of the signal can be separated into an average Ramsey carrier and a slowly varying complex envelope. 
For the constant-detuning case considered in this derivation, we define
\[
\bar{\omega}=\frac{\omega_1+\omega_2}{2},
\qquad
\phi=(\omega_1-\omega_2)t,
\]
where \(\phi\) is the accumulated relative phase between the two Ramsey
components. The weighted sum of the two Ramsey components can then be written as
\begin{equation}
A_1\cos(\omega_1 t)+A_2\cos(\omega_2 t)
=
\operatorname{Re}\!\left[
e^{i\bar{\omega}t}
\left(
A_1e^{i\phi/2}+A_2e^{-i\phi/2}
\right)
\right].
\end{equation}

The slowly varying envelope is therefore
\begin{equation}
C(\phi)
\equiv
A_1 e^{i\phi/2}+A_2 e^{-i\phi/2}
=
\cos\left(\frac{\phi}{2}\right)
+
i\epsilon\sin\left(\frac{\phi}{2}\right),
\label{eq:SM_complex_amplitude}
\end{equation}
where
\begin{equation}
\epsilon=A_1-A_2 .
\end{equation}
Here \(\epsilon\) is kept as a signed imbalance.  Quantities such as visibility
and gain depend on \(|\epsilon|\), while the sign of \(\epsilon\) determines
the direction of the projected phase response.

\begin{figure}[h]
    \centering
    \includegraphics[width=\linewidth]{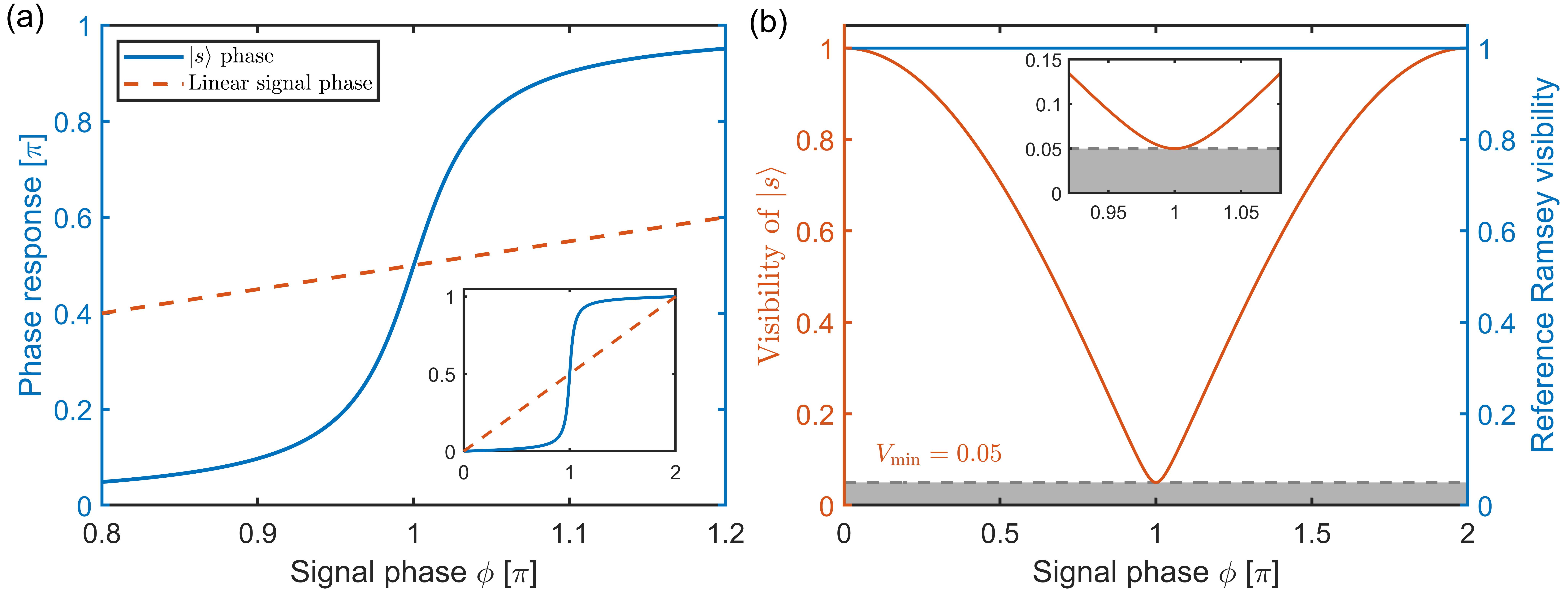}
\caption{
\textbf{Comparison between the one-path Ramsey reference and the geometric
phase response.}
\textbf{(a)} Phase response as a function of the accumulated signal phase
$\phi$. The dashed orange curve shows the linear signal-phase response
expected from the one-path Ramsey reference. The solid blue
curve shows the phase of the projected state $|s\rangle$, which exhibits a
rapid geometric phase transition near $\phi=\pi$. The inset shows the
corresponding behavior over the full $0$ to $2\pi$ range.
\textbf{(b)} Visibility response for the same parameter sweep. The visibility
of the projected state $|s\rangle$ reaches the minimum $V_{\min}=0.05$ near
$\phi=\pi$, as highlighted by the gray region and the inset, whereas the
reference Ramsey visibility remains unity. These curves correspond to
representative cuts of the phase and visibility landscapes in main Fig.\,2:
the dashed phase and reference Ramsey visibility illustrate the standard
two-level response, while the projected state $|s\rangle$ shows the geometric
phase-amplified response for
$|\epsilon|=|A_1-A_2|=0.05$.
}
    \label{fig:S_phase_visibility_cuts}
\end{figure}

\begin{figure}[h]
    \centering
    \includegraphics[width=\linewidth]{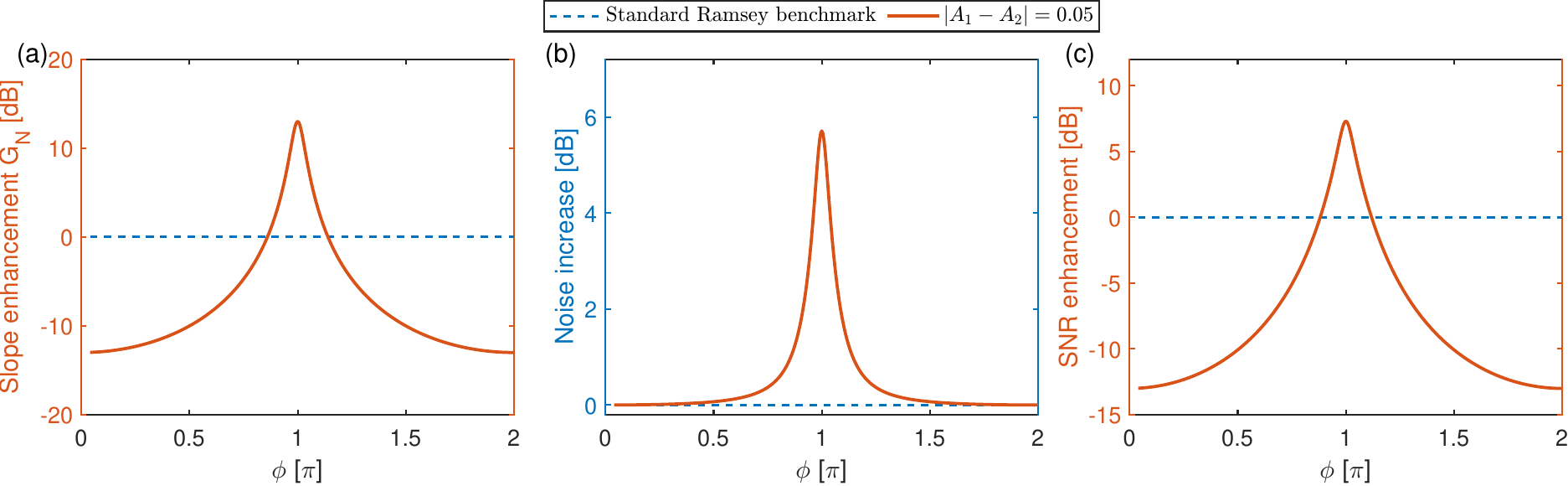}
    \caption{
    \textbf{Sensitivity enhancement along the geometric-response cut.}
    Line cuts of main Fig.\,3 for the representative near-balanced case
    \(|\epsilon|=|A_1-A_2|=0.05\). The orange curve shows the geometric-response cut, while the
blue dashed line marks the one-path Ramsey benchmark.
\textbf{(a)} Phase-response slope enhancement $G_N$
relative to this benchmark.
    \textbf{(b)} Visibility-induced phase-noise increase associated with
    the reduced projected visibility near \(\phi=\pi\).
    \textbf{(c)} Net SNR enhancement after accounting for both the slope gain
    and the visibility-induced noise penalty. The positive region near
    \(\phi=\pi\) identifies the finite operating range where the geometric response improves
the input-referred sensitivity. Calculations use \(N=3\times10^3\) and
    \(\xi_{\rm tech}=0.1\,{\rm rad}\).
    }
    \label{fig:S_sensitivity_cut}
\end{figure}

The derivation applies beyond the V-type schematic in Fig.\,1. A ladder or \(\Lambda\)-type three-level system gives the same effective envelope in Eq.\,(\ref{eq:SM_complex_amplitude}), after choosing the appropriate transition-phase sign convention, provided that the pulse sequence creates two coherent signal-accumulating pathways and the final readout projects them onto a common state. The physical level ordering changes the implementation of the pulses and detunings while preserving the projected-interference form used here.

Writing the complex envelope as
\begin{equation}
C(\phi)=V_s(\phi)e^{i\Phi(\phi)},
\end{equation}
the projected shared-state population becomes
\begin{equation}
P_s(t)
=
\frac{1}{2}
\left[
1
-
V_s(\phi)
\cos\left(\bar{\omega}t+\Phi(\phi)\right)
\right].
\label{eq:SM_projected_signal}
\end{equation}
Thus the effective visibility is
\begin{equation}
V_s(\phi)
=
\sqrt{
\cos^2\left(\frac{\phi}{2}\right)
+
\epsilon^2\sin^2\left(\frac{\phi}{2}\right)
},
\label{eq:SM_visibility}
\end{equation}
and the projected readout phase is
\begin{equation}
\Phi(\phi)
=
\arctan
\left[
\frac{
\epsilon \sin\!\left(\phi/2\right)
}{
\cos\!\left(\phi/2\right)
}
\right].
\label{eq:SM_phase}
\end{equation}

\section{Visibility minimum and normalized phase gain}
\label{sec:SM_gain_visibility}

We now use the effective visibility and projected phase derived above to
identify the visibility minimum and the local phase-response gain.  From
Eq.\,\eqref{eq:SM_visibility},
\begin{equation}
V_s^2(\phi)
=
1-
\left(1-\epsilon^2\right)
\sin^2\left(\frac{\phi}{2}\right).
\label{eq:SM_visibility_squared}
\end{equation}
The visibility therefore reaches its minimum at
\begin{equation}
\phi=(2m+1)\pi,
\qquad
V_{s,\min}=|\epsilon|=|A_1-A_2|.
\label{eq:SM_visibility_minimum}
\end{equation}
For a nearly balanced preparation, \(|A_1-A_2|\ll1\), the projected fringe
visibility becomes small near \(\phi=\pi\).

The local phase response follows by differentiating
Eq.\,\eqref{eq:SM_phase}:
\begin{equation}
\frac{\partial \Phi}{\partial \phi}
=
\frac{\epsilon}
{2\left[
\cos^2\left(\frac{\phi}{2}\right)
+
\epsilon^2\sin^2\left(\frac{\phi}{2}\right)
\right]}
=
\frac{\epsilon}{2V_s^2(\phi)} .
\label{eq:SM_phase_slope}
\end{equation}
The sign is set by the signed pathway imbalance \(\epsilon\), while the
response magnitude depends on \(|\epsilon|\).

We define the normalized phase-response gain relative to the ordinary
half-phase Ramsey response as
\begin{equation}
G_N(\phi)
\equiv
2\left|
\frac{\partial \Phi}{\partial \phi}
\right|
=
\frac{|\epsilon|}{V_s^2(\phi)} .
\label{eq:SM_normalized_gain}
\end{equation}
At the visibility minimum,
\begin{equation}
G_N(\pi)=\frac{1}{|\epsilon|}.
\label{eq:SM_gain_at_transition}
\end{equation}
Thus reducing the pathway imbalance increases the local phase-response gain,
but simultaneously lowers the visibility at the transition.  This
visibility--gain tradeoff underlies the metrological behavior discussed in the
main text.

\section{Representative phase and visibility cuts}

\noindent The expressions above show that the projected signal can be written as an
effective Ramsey fringe with a phase response $\Phi$ and visibility $V_s$.
To illustrate the physical meaning of these quantities, Fig.\,\ref{fig:S_phase_visibility_cuts} shows a
representative nearly balanced case with $|\epsilon|=|A_1-A_2|=0.05$. In this regime, the projected phase exhibits a rapid noncyclic variation near the geodesic-closure transition at \(\phi=\pi\), whereas the differential phase of the underlying two-level components varies smoothly. The rapid phase response is accompanied by a reduction in projected
visibility, while the reference Ramsey visibility remains unity.

\section{Representative cut of the sensitivity landscape}
\label{sec:SM_sensitivity_cut}

The metrological performance depends on both the slope enhancement and the
visibility-induced phase-noise cost. Near the noncyclic transition, the
enhanced phase response is accompanied by reduced projected visibility.
The sensitivity landscape in the main text combines these two effects to
determine the net SNR enhancement.

Figure\,\ref{fig:S_sensitivity_cut} shows the corresponding line cut for the
representative near-balanced case \(|\epsilon|=|A_1-A_2|=0.05\). The slope
enhancement peaks near \(\phi=\pi\), where the visibility reaches its minimum.
The visibility-induced noise increase partially offsets this gain, so the net
SNR enhancement is positive only within a finite operating region around the
transition. This cut illustrates the same gain--visibility tradeoff shown in
the full landscape of main Fig.\,3.

\section{Atom-number and technical-noise dependence of the projected SNR enhancement}
\label{sec:SM_noise_number_landscape}

The main text evaluates the geometric response at a representative operating
point and uses a finite-window average over the offset-shifted sensitivity
window. Here we complement that analysis by showing how the locally optimized
projected SNR enhancement depends on two experimental parameters: the atom
number \(N\) and the technical-noise level
\(\xi_{\rm tech}\). For this calculation, we use the same pathway imbalance as in the main text,
$|\epsilon|=|A_1-A_2|=0.05$,
corresponding to a visibility minimum $V_{s,\min}=0.05$
and a peak normalized phase gain
$G_N^{\max}=1/|\epsilon|=20$. For each pair \((N,\xi_{\rm tech})\), the SNR
enhancement is maximized over the accumulated phase and quoted relative to a
standard Ramsey benchmark with the same \(N\) and \(\xi_{\rm tech}\).

\begin{figure}[h]
    \centering
    \includegraphics[width=\linewidth]{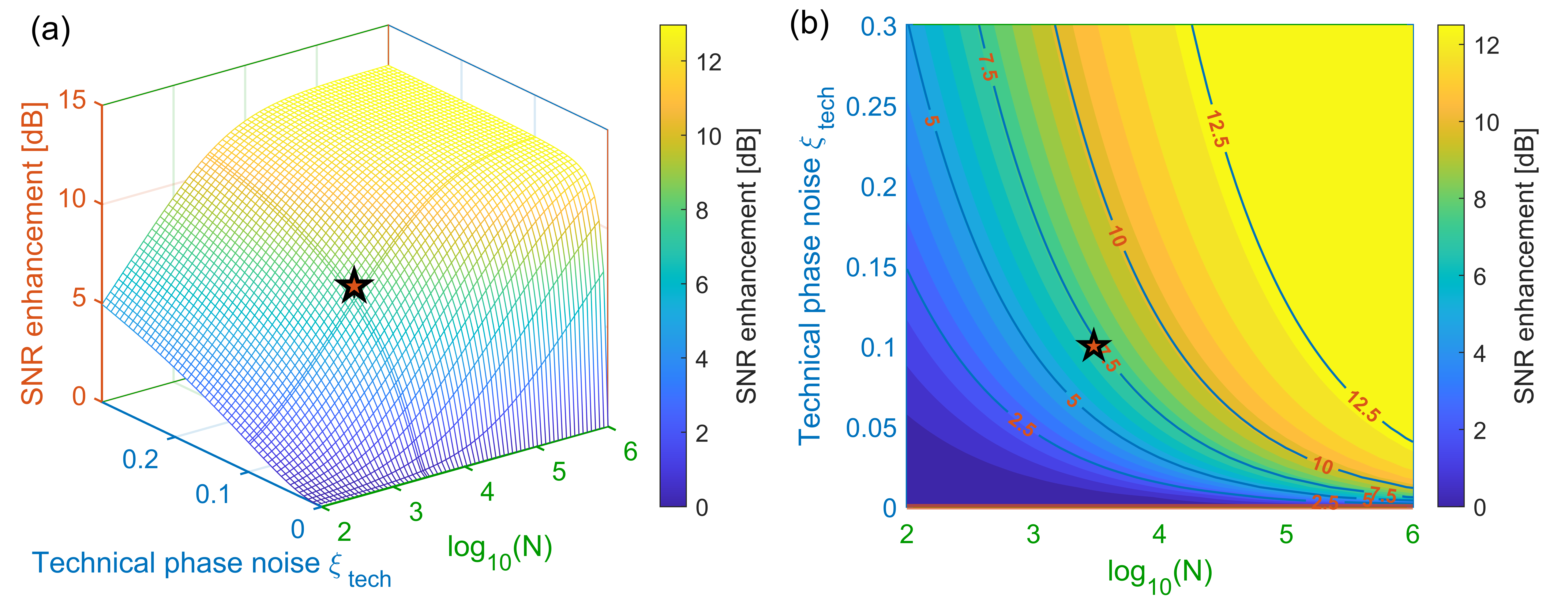}
\caption{
\textbf{Projected SNR enhancement versus atom number and technical
phase noise.}
(a) Surface plot of the locally optimized SNR enhancement as a function of
atom number \(N\) and additional technical phase noise
\(\xi_{\rm tech}\).
(b) Corresponding contour plot. The calculation uses
\( |\epsilon|=|A_1-A_2|=0.05 \), corresponding to
\(V_{s,\min}=0.05\) and \(G_N^{\max}=20\). For each pair
\((N,\xi_{\rm tech})\), the enhancement is maximized over the accumulated
phase and quoted in dB relative to the one-path Ramsey benchmark under the
same \(N\) and \(\xi_{\rm tech}\), using the convention
\(10\log_{10}({\rm SNR}_{\rm geo}/{\rm SNR}_{\rm ref})\).
The red line at \(\xi_{\rm tech}=0\) indicates the
projection-noise-limited static-phase case, where the maximum projected
enhancement is \(0\,{\rm dB}\) because the geometric slope gain is
compensated by the visibility-induced phase-noise penalty. The enhancement
increases as the technical phase noise becomes appreciable, because the
enhanced phase response suppresses its contribution when referred back to
the inferred signal phase. The star marks the representative parameter set
used in the main text, \(N=3\times10^{3}\) and
\(\xi_{\rm tech}=0.1\,{\rm rad}\). At this point, the locally optimized
enhancement is about \(7.3\,{\rm dB}\), while the finite-window effective
enhancement used in Fig.\,4(d) is about \(7.0\,{\rm dB}\).
}
    \label{fig:S_noise_number_landscape}
\end{figure}

The corresponding local SNR ratio, including both the geometric gain and the visibility-induced noise penalty, can then be written as
\begin{equation}
R_{\rm SNR}(\phi;N,\xi_{\rm tech})
=
G_N(\phi)
\frac{
\sqrt{1/N+\xi_{\rm tech}^2}
}{
\sqrt{1/[V_s^2(\phi)N]+\xi_{\rm tech}^2}
}.
\label{eq:SM_local_SNR_ratio}
\end{equation}
The quantity plotted in Fig.~\ref{fig:S_noise_number_landscape} is the
locally optimized SNR enhancement,
\begin{equation}
\mathcal{G}_{\rm SNR}(N,\xi_{\rm tech})
=
10\log_{10}
\left[
\max\nolimits_{\phi} R_{\rm SNR}(\phi;N,\xi_{\rm tech})
\right],
\label{eq:SM_SNR_landscape}
\end{equation}
where the convention is \(10\log_{10}({\rm SNR}_{\rm geo}/{\rm SNR}_{\rm ref})\).

When \(\xi_{\rm tech}=0\), the maximum projected enhancement is
\(0\,{\rm dB}\) in this static-phase noise model. In the purely
projection-noise-limited case, the geometric slope gain is compensated by the
visibility-induced increase in phase noise. When additional technical noise is appreciable, the enhanced phase-response slope reduces its input-referred contribution to the inferred signal phase.

Figure\,\ref{fig:S_noise_number_landscape} isolates the gain--visibility
tradeoff within the noise model of Eq.\,(\ref{eq:SM_local_SNR_ratio}) and maps the dependence of the
locally optimized SNR enhancement on atom number and additional technical
noise. Because the figure shows a relative SNR ratio, the absolute
$N^{-1/2}$ improvement with atom number is implicit. At the transition,
$G_N V_s=1$, so the input-referred projection-noise term remains
$1/\sqrt{N}$ while the technical-noise contribution is reduced to
$\xi_{\rm tech}/G_N$. For fixed $\xi_{\rm tech}$, the crossover to the
technical-noise-dominated regime therefore occurs at an atom number larger
by approximately $G_N^2$ than for standard Ramsey, extending the
projection-noise-dominated improvement to larger sensor numbers.
System-specific effects such as decoherence, density shifts, collisions,
imperfect state preparation, and other technical-noise mechanisms remain
implementation dependent.


\section{Finite-window SNR enhancement used for the phase projection}

The phase projection in Fig.\,4(d) uses the local response near the final
offset-controlled operating point, rather than an average over the full phase
trajectory during the Ramsey evolution. In a Ramsey shot, the signal phase is
accumulated during the interrogation and read out at the end; the relevant
response is therefore determined by the final accumulated phase. The initial
offset places this final phase near the noncyclic branch region.

The enhanced response is local: both the normalized phase gain $G_N(\phi)$
and the projected visibility $V_s(\phi)$ vary near the branch region. To avoid using only the peak tangent gain, we evaluate the local SNR
enhancement $R_{\rm SNR}(\phi;N,\xi_{\rm tech})$ defined in Eq.\,(\ref{eq:SM_local_SNR_ratio}) at
each final phase in the shaded operating window $W$, and then combine these
local enhancements in inverse-variance form. The effective enhancement used in the finite-window
phase projection is
\begin{equation}
E_{\rm eff}
=
\sqrt{
\left\langle
R_{\rm SNR}^2(\phi;N,\xi_{\rm tech})
\right\rangle_W
},
\label{eq:SM_finite_window_enhancement}
\end{equation}
where $\langle\cdots\rangle_W$ denotes a uniform average over the shaded
window of final phases near the high-slope response.

The inverse-variance average in Eq.\,\eqref{eq:SM_finite_window_enhancement}
follows the standard Fisher-information view of combining independent local
estimates\,[S1], consistent with the broader estimation framework used
in quantum metrology\,[S2,S3].


This window-averaged treatment gives a conservative estimate compared with
using the peak local response at the exact branch point. It is most directly
applicable to phase signals or burst-induced integrated phase shifts that can
be sampled within the finite high-slope operating window. For quasi-static
frequency shifts, the signal accumulation time must also be included, as
discussed in the main text.

\section{Trigonometric derivation of the projected phase and visibility}
\label{sec:SM_phase_visibility_derivation}

The complex-amplitude form used above gives the most compact route to
\(V_s(\phi)\) and \(\Phi(\phi)\). For readers who prefer to avoid this
complex-valued representation, we give here an equivalent trigonometric
derivation. This also provides a direct check of the projected Ramsey signal
used in the main text. The shared-state population after the final readout
pulses is
\begin{equation}
P_s(t)
=
\frac{A_1}{2}\left[1-\cos(\omega_1 t)\right]
+
\frac{A_2}{2}\left[1-\cos(\omega_2 t)\right],
\label{eq:SM_trig_start}
\end{equation}
where \(A_1+A_2=1\). We define
\begin{equation*}
\bar{\omega}=\frac{\omega_1+\omega_2}{2},
\qquad
\Delta\omega=\omega_1-\omega_2,
\qquad
\phi=\Delta\omega t,
\end{equation*}
so that
\begin{equation*}
\omega_1=\bar{\omega}+\frac{\Delta\omega}{2},
\qquad
\omega_2=\bar{\omega}-\frac{\Delta\omega}{2}.
\end{equation*}
The signed pathway imbalance is
\begin{equation*}
\epsilon=A_1-A_2 .
\end{equation*}

Expanding Eq.\,\eqref{eq:SM_trig_start} gives
\begin{equation*}
P_s(t)
=
\frac{1}{2}
-
\frac{1}{2}
\left[
A_1\cos(\omega_1 t)+A_2\cos(\omega_2 t)
\right].
\end{equation*}
Using
\begin{align*}
\cos(\omega_1 t)
&=
\cos(\bar{\omega}t)\cos\left(\frac{\phi}{2}\right)
-
\sin(\bar{\omega}t)\sin\left(\frac{\phi}{2}\right),\\
\cos(\omega_2 t)
&=
\cos(\bar{\omega}t)\cos\left(\frac{\phi}{2}\right)
+
\sin(\bar{\omega}t)\sin\left(\frac{\phi}{2}\right),
\end{align*}
we obtain
\begin{align*}
A_1\cos(\omega_1 t)+A_2\cos(\omega_2 t)
&=
(A_1+A_2)\cos(\bar{\omega}t)\cos\left(\frac{\phi}{2}\right)\\
&\quad+
(A_2-A_1)\sin(\bar{\omega}t)\sin\left(\frac{\phi}{2}\right)\\
&=
\cos(\bar{\omega}t)\cos\left(\frac{\phi}{2}\right)
-
\epsilon\sin(\bar{\omega}t)\sin\left(\frac{\phi}{2}\right).
\end{align*}

We now write this expression as a single effective Ramsey fringe,
\begin{equation*}
A_1\cos(\omega_1 t)+A_2\cos(\omega_2 t)
=
V_s(\phi)\cos\!\left[\bar{\omega}t+\Phi(\phi)\right].
\end{equation*}
Expanding the right-hand side gives
\begin{equation*}
V_s\cos\left(\bar{\omega}t+\Phi\right)
=
V_s\cos\Phi\,\cos(\bar{\omega}t)
-
V_s\sin\Phi\,\sin(\bar{\omega}t).
\end{equation*}
Matching the coefficients of \(\cos(\bar{\omega}t)\) and
\(\sin(\bar{\omega}t)\) yields
\begin{equation*}
V_s(\phi)\cos\Phi(\phi)
=
\cos\left(\frac{\phi}{2}\right),
\qquad
V_s(\phi)\sin\Phi(\phi)
=
\epsilon\sin\left(\frac{\phi}{2}\right).
\end{equation*}
Therefore,
\begin{equation*}
V_s(\phi)
=
\sqrt{
\cos^2\left(\frac{\phi}{2}\right)
+
\epsilon^2
\sin^2\left(\frac{\phi}{2}\right)
},
\end{equation*}
and
\begin{equation*}
\Phi(\phi)
=
\arctan
\left[
\frac{
\epsilon \sin\!\left(\phi/2\right)
}{
\cos\!\left(\phi/2\right)}\right].
\end{equation*}
Here $\Phi(\phi)$ is taken continuously across $\phi=\pi$.

Substituting these expressions back into \(P_s(t)\) yields
\begin{equation}
P_s(t)
=
\frac{1}{2}
\left[
1
-
V_s(\phi)
\cos\left(\bar{\omega}t+\Phi(\phi)\right)
\right].
\label{eq:SM_trig_projected_signal}
\end{equation}
Thus, near \(\phi=\pi\), the visibility reaches the finite minimum
\(V_{s,\min}=|\epsilon|\), while the projected readout phase changes rapidly.

\section*{Supplementary References}

\noindent\hypertarget{Sref:Kay1993}{[S1]}
S. M. Kay, \textit{Fundamentals of Statistical Signal Processing, Volume I: Estimation Theory}
(Prentice Hall, Upper Saddle River, NJ, 1993).

\noindent\hypertarget{Sref:BraunsteinCaves1994}{[S2]}
S. L. Braunstein and C. M. Caves, ``Statistical distance and the geometry of quantum states,''
\textit{Phys. Rev. Lett.} \textbf{72}, 3439--3443 (1994).

\noindent\hypertarget{Sref:Paris2009QuantumEstimation}{[S3]}
M. G. A. Paris, ``Quantum estimation for quantum technology,''
\textit{Int. J. Quantum Inf.} \textbf{7}, 125--137 (2009).

\end{document}